\documentstyle[prl,aps,epsfig,twocolumn]{revtex}
\begin{document}
\draft
\flushbottom
\twocolumn[
\hsize\textwidth\columnwidth\hsize\csname @twocolumnfalse\endcsname

\title{Microscopic model of superconductivity in carbon nanotubes}
\author{J. Gonz\'alez   \\}
\address{
        Instituto de Estructura de la Materia.
        Consejo Superior de Investigaciones Cient{\'\i}ficas.
        Serrano 123, 28006 Madrid. Spain.}

\date{\today}
\maketitle
\begin{abstract}
\widetext
We propose the model of a manifold of one-dimensional interacting 
electron systems to account for the superconductivity observed 
in ropes of nanotubes. We rely on the strong suppression of 
single-particle hopping between neighboring nanotubes in 
a disordered rope and conclude that the tunnelling takes place 
in pairs of electrons, which are formed within each nanotube 
due to the existence of large superconducting correlations. 
Our estimate of the transition temperature is consistent 
with the values that have been measured experimentally in 
ropes with about 100 metallic nanotubes.

\end{abstract}

]

\narrowtext 
\tightenlines


Carbon nanotubes have nowadays a great potential for 
technological applications in devices at the nanometer
scale. This makes very important the precise knowledge
of their electronic properties, taken as individual
structures as well as when packed in the form 
of ropes. The first step of discerning their metallic 
or insulating behavior has been completed, as the 
results obtained from theoretical 
considerations\cite{saito} 
have been confirmed experimentally\cite{wild}. 
It turns out that the low-energy spectrum of a carbon 
nanotube may be gapless or not
depending on the particular wrapping of the graphene
sheet to form the tubular structure.

The electron interactions are also known to modify
significantly the transport properties of the 
nanotubes\cite{exp,yao}. 
In the case of metallic single-walled nanotubes, the
one-dimensional (1D) character of the system leads to a strong
correlation among the electrons, making the so-called 
Luttinger liquid\cite{emery,lutt1,lutt2,sch} 
the most appropriate paradigm 
to describe the electronic properties\cite{bal,eg,kane}. 
Evidence of Luttinger liquid
behavior has been found in measurements of the tunnelling
conductance in ropes\cite{exp} and in individual 
single-walled nanotubes\cite{yao}. 

Experiments have been also carried out to probe 
superconducting correlations in the carbon 
nanotubes\cite{kas,marcus}. By
suspending them between superconducting contacts, large 
supercurrents have been measured in samples almost made of
a single-walled nanotube and in ropes\cite{kas}. 
More recently, superconducting properties have been 
measured in ropes suspended between nonsuperconducting,
good metallic contacts\cite{sup}. Evidence of 
superconducting fluctuations has been also obtained 
in individual single-walled nanotubes\cite{chi}. 
In the experiments presented in Ref. \onlinecite{sup}, 
a drop by two orders of magnitude in the resistance
has been found, down to the minimum value consistent 
with the number of conducting channels in the rope. 
This feature, together with its suppression under a 
suitably high magnetic field, shows the existence of
superconductivity inherent to the ropes of 
nanotubes\cite{sup}. The value of the 
transition temperature varies from one sample to other, 
being below 1 K in the two where the resistance drop 
has been measured. 

In this Letter we develop a model to account for the 
superconductivity intrinsic to the ropes of nanotubes. 
The problem that faces any model trying to describe 
such effect is twofold. In the first place, 
it is known that the single-particle hopping between
neighboring nanotubes in a rope is strongly 
suppressed\cite{mkm}. This is because, in general, 
the different helical structure of the
nanotubes leads to the misalignement of their lattices
and, therefore, to the difficulty of conserving 
momentum for an electron hopping from one nanotube
to the other. On the other hand, the effect of 
superconductivity cannot rely exclusively on the 
properties of the individual nanotubes, since any
correlation in a 1D system can only
develop a divergence at zero temperature\cite{sch}. 

The experiments on the proximity effect
of Ref. \onlinecite{kas} show anyhow 
the existence of sensible superconducting fluctuations
in carbon nanotubes. The precise value of the 
critical supercurrents found there can be only 
explained by the presence of a short-range attractive 
interaction coming from the coupling to the elastic 
modes of the nanotube\cite{prl}. The point is that,
as long as Cooper pairs are formed locally with zero
total momentum, they may overcome the special
difficulty that single electrons find to tunnel
between neighboring nanotubes.

According to the work of Ref. \onlinecite{mkm}, the
amplitude $t_T$ for tunnelling between nanotubes with 
the same chirality and orientation is of the order
$t_T \sim 0.01$ eV. In a disordered rope, the
misalignement of the lattices of neighboring nanotubes
introduces in general an additional suppression of the
hopping amplitude by a factor $ \sim
\exp (- R a_0 (\delta k)^2/4) $, where $R$ is the radius
of the tube, $\delta k$ is related to the mismatch of
the Fermi points, and $a_0$ is a parameter of the order 
$\sim 0.5$ \AA \cite{mkm}.
For a typical nanotube radius $R = 7$ \AA, this factor
is $ \sim 0.005 $. 

The tunnelling amplitudes have to be compared with 
the energy scale at which the metallic nanotubes behave 
as 1D objects. 
This is the scale $E_c$ below which the gapless
linear subbands dominate the physical properties, 
and it can be estimated as $E_c \sim 0.1$ eV. 
The probability $\lambda_2$ of tunnelling of a Cooper 
pair, given by the square of $t_T$ in units of $E_c$,
has a relative weight of the order $\sim 0.01$. 
On the other hand, the probability of
tunnelling of a single electron has a relative weight
of the order $\sim 0.0005$. This means that pair hopping
is the dominant process for tunnelling between neighboring
nanotubes in a disordered rope.

Prior to considering pair hopping, the rope can be described
at low energies by a model in which each nanotube is treated 
as a 1D system, whose charge can interact
with the charge in the other nanotubes of the rope. 
The hamiltonian for this model, including a collection of
metallic nanotubes $a = 1, \ldots n$ with linear
branches of different chirality $i = +,-$, can be written
in terms of the respective density operators
$\rho_{a i \sigma }$ \cite{rem}:
\begin{eqnarray}
H_1  & = &    \frac{1}{2} v_F \int_{-k_c}^{k_c} dk 
 \sum_{a i \sigma }  : \rho_{a i \sigma} (k)
             \rho_{a i \sigma} (-k)  :   \nonumber      \\
  &  &    + \frac{1}{2}  \int_{-k_c}^{k_c} dk \; 
      \sum_{a i \sigma } \rho_{a i \sigma} (k) \; 
  \sum_{b j \sigma'  }  V_{ab}(k)  \;
                      \rho_{b j \sigma'} (-k)
\label{ham}
\end{eqnarray}
where 
$k_c = E_c/v_F$.

For the potential between different nanotubes 
$V_{ab}$ we take the Coulomb interaction     
$V_C (k) = e^2 /(4\pi^2)
      \log |(k_c + k)/ k|$ \cite{wang}, which remains 
long-ranged in one spatial dimension\cite{grab}. 
According to the results of Ref. \onlinecite{prl}, the 
interaction potential $V_{aa}$ within each nanotube includes
moreover the effective short-range attraction coming from
the coupling to the elastic modes, so that
$V_{aa}(k) = e^2 /(4\pi^2) \log |(k_c + k)/ k| - g/(2\pi )$.
The strength $g$ of the attractive interaction
is inversely proportional to the mass of the atoms and
directly proportional to $(t'/v_s)^2$, where $t'$ is the
modulation of the lattice hopping and $v_s$ is the speed
of sound\cite{loss}. A rough estimate for a carbon 
nanotube gives $g/v_F \sim 0.2$ .

Terms which couple the spin densities have been neglected
in writing the hamiltonian (\ref{ham}).
These are backscattering (BS) interactions, which arise as 
a remnant of the Coulomb interaction at short distances.
Being local in the nanotube lattice, their couplings are 
reduced by a relative factor of the order $\sim 0.1 a/R$, 
where $a$ is the nanotube lattice spacing\cite{bal,eg,kane}. 
These terms are marginally relevant in the renormalization 
group sense. This means that they have greater strength as 
the model is scaled to smaller energies, but the rate of
increase starts being quadratic in the own couplings. 
Thus, the theory has to be scaled to extremely low 
energies, many orders of magnitude below $E_c$, to have the 
BS couplings comparable to the couplings in 
(\ref{ham}) \cite{eg}.

The bundle of 1D electron systems coupled only
by charge interactions ressembles the system proposed in
Ref. \onlinecite{slide} for the description of the sliding
Luttinger liquid. Here the Coulomb interaction couples each 
of the nanotubes to all the others in the bundle.
We recall that, for the samples of the experiments reported
in Ref. \onlinecite{sup},
the number of metallic nanotubes is very
large, of the order $n \sim 100$. Then, the coupling
of the charge in the different nanotubes leads to a
significant reduction of the effective repulsive
interaction. On the other hand, the effect of the intra-tube
attractive interaction does not depend on the number of
nanotubes and, in certain regime, it may dominate over
the Coulomb interaction.

The above point can be checked by looking at the
correlators of the model governed by $H_1$. These can be
computed exactly by means of bosonization techniques.
For instance, the propagator $D^{(0)}(x,t)$ for the Cooper 
pairs within each nanotube factorizes into the different 
interaction channels, taking the form 
\begin{eqnarray}
 D^{(0)}(x,t) & \equiv &   
          \langle \Psi^{+}_{a + \uparrow} (x,t)
   \Psi^{+}_{a - \downarrow} (x,t) \Psi_{a + \uparrow} (0,0)   
     \Psi_{a - \downarrow} (0,0)  \rangle  \nonumber  \\ 
      &  =  &   C(x,t) \prod_{1}^{n-1} N(x,t)
               \prod_{1}^{3n} F(x,t)
\label{coop}
\end{eqnarray}

The first factor in Eq. (\ref{coop}) corresponds to the 
contribution of the total charge density, which is given 
at zero temperature by the expression
\begin{equation}
C(x,t) = \exp \left( -\frac{1}{2n} \int_{0}^{k_c}
     dk \frac{1}  {\mu (k)\: k}
      \left(1 - \cos (kx) \: \cos (\tilde{v}_F kt) \right)
            \right)
\label{prop}
\end{equation}
where $\mu (k) = 1/\sqrt{1 + 8n V_{C}(k)/v_F - 4g/(\pi v_F)}$ 
and $\tilde{v}_F (k) = v_F/\mu (k)$.
The factors $N(x,t)$ correspond to the rest of the charge
channels, and they have a form similar to (\ref{prop}) but
with $\mu  = 1/\sqrt{1 - 4g/(\pi v_F)}$ instead of 
$\mu (k)$ and $\tilde{v}_F  = v_F/\mu $ instead of
$\tilde{v}_F (k)$. The factors $F(x,t)$ correspond to the
noninteracting channels with $\mu = 1$.
The computation can be extended to the case of 
temperature $T \neq 0$, just by inserting the factor 
$1 + 2/[\exp (\tilde{v}_F |k|/T) - 1]$ in the integrand of 
expressions such as (\ref{prop}) \cite{emery}.

\begin{figure}[H]

\epsfig{file=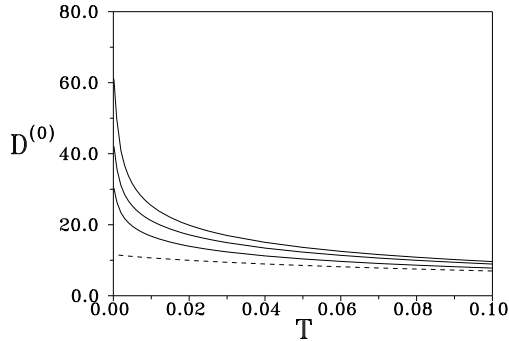, height=7cm, width=7cm,
         bbllx=0, bblly=0, bburx=600, bbury=600,
         angle=270}

\caption{Plot of the propagator $\widetilde{D}^{(0)}$ at
zero frequency and momentum versus $T/E_c$, for 
$2 e^2/(\pi^2 v_F) = 1.0$ . The dashed line 
corresponds to the case $n = 1$ and $g = 0$, and the 
solid lines to $n = 100$ and the respective values 
(from top to bottom) $4g/(\pi v_F) = 0.75, 0.5, 0$ .}
\label{one}
\end{figure}

The couplings that produce the most accurate fit of the
critical supercurrents reported in Ref. \onlinecite{kas} 
are $2 e^2/(\pi^2 v_F) \approx 1.0$ and $4g/(\pi v_F) 
\approx 0.6$ \cite{prl}. For $n \sim 100$, the strong
reduction of the Coulomb interaction implies that the 
effect of the intra-tube attractive interaction prevails
in the system. In the $g$-ology 
description\cite{lutt1}, the model is in the regime with 
short-range attractive coupling $g_2 < 0$, where the singlet
superconductor response is enhanced over the 
charge-density-wave response. 
The temperature dependence of the Fourier
transform $\widetilde{D}^{(0)}$ of the propagator at zero 
frequency and momentum is represented in Fig. \ref{one}.
The enhancement of the propagator is the signal that
large superconducting correlations exist in the individual
nanotubes at low temperatures.

The above considerations are pertinent
to the system at half-filling, in which the linear subbands
cross at the Fermi level. When the nanotubes are slightly doped,  
the shift of the Fermi level gives rise to four Fermi points. 
The Cooper pairs have then the possibility to resonate between 
the outer and the inner gapless subbands. 
Anyhow, as long as the repulsive interactions
mediating these processes are reduced by a relative 
factor as small as that of
the BS interactions\cite{bal,eg}, the $s$-wave pairing is 
favored over other channels with different symmetry.


The tunnelling of Cooper pairs between the nanotubes can 
be taken into account by modifying the hamiltonian 
(\ref{ham}) with the additional term
\begin{eqnarray}
 H_2 =  \sum_{\langle a,b \rangle}   ( \lambda_2 )_{ab} 
  \int_{-k_c}^{k_c} dk   \int_{-k_c}^{k_c} dp
     \int_{-k_c}^{k_c} dp'
              \;\;\;\;\;\;\;\;\;\;\;\;\;\;\;
                  \;\;\;\;\;&   &      \nonumber       \\
  \left(  \Psi^{+}_{a i \uparrow}(k+p)
          \Psi^{+}_{a -i \downarrow}(-p)
           \Psi_{b j \uparrow}(k+p')
           \Psi_{b -j \downarrow}(-p')
              + {\rm h. c.} \right) &  &
\label{ham2}
\end{eqnarray}
where $\Psi^{+}_{a i \sigma}$ is the electron
operator and
the sum runs over all pairs $\langle a,b \rangle$ of
nearest-neighbor metallic nanotubes.

The term (\ref{ham2}) is a relevant perturbation
from the renormalization group point of view.
However, the anomalous scaling dimensions of the relevant
perturbations turn out to be in general rather small. 
They can be computed in the boson representation and, 
in the case of the term (\ref{ham2}), the result is
$\gamma_2 = 2[1/(4n\mu (k_0)) + (n-1)/(4n\mu ) - 1/4]$.
$k_0$ represents some effective value, which does 
not affect significantly the estimate. For the above mentioned 
couplings, we obtain $\gamma_2 \approx - 0.07$. Thus, even 
four orders of magnitude below $E_c$, we observe that 
$\lambda_2$ is not enhanced by more than a factor
$(10^{-4})^{\gamma_2} \sim 2$.

\begin{figure}
\begin{center}
\mbox{\epsfxsize 8.5cm \epsfbox{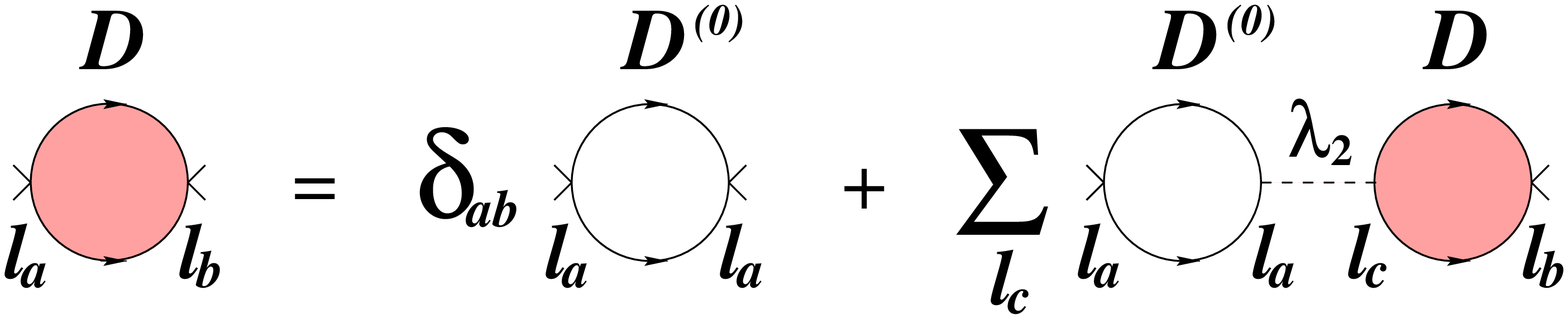}}
\end{center}
\caption{Self-consistent diagrammatic equation for the 
propagator $D$ of Cooper pairs along the rope.}
\label{two}
\end{figure}

The proposed model has a
superconducting instability at some finite temperature,
provided that the Cooper pairs are able to percolate   
in the transverse directions of the rope.
This is the case of the superconducting ropes
of Ref. \onlinecite{sup}, although to meet such
experimental condition one has to find the appropriate      
sample out of a large number of them\cite{com}.
Assuming the coherence in the hopping of pairs in the 
transverse directions,
we may write the propagator of the Cooper pair 
from a metallic nanotube $a$ to
another $b$ as a function of the distance $x$ along the
rope and the positions $l_a$ and $l_b$ of the nanotubes
in the transverse section of the rope, 
$D (l_a,l_b;x,t)$. This object can be related to the
propagator $D^{(0)}$ along each nanotube through the
self-consistent diagrammatic equation in Fig. \ref{two},
which takes into account the dominant terms in powers of
$\lambda_2$.

By introducing the Fourier transform with respect to 
$(x,t)$ as well as in the $l_a$ variables\cite{rem2}, 
the equation in Fig. \ref{two} can be written for the 
Fourier transformed propagator $\widetilde{D}$ in the 
form
\begin{equation}
\widetilde{D} (q;k,\omega_k) =
         \widetilde{D}^{(0)} (k,\omega_k) +
    \widetilde{D}^{(0)} (k,\omega_k) \lambda_2 (q) 
                       \widetilde{D} (q;k,\omega_k)
\label{fourier}
\end{equation}

We are interested in the propagation of the Cooper pairs
from a metallic nanotube to the rest in the rope, which
is given by $\widetilde{D} (0;0,0) =
    \sum_{l_b} \int dx \int dt \;  D(l_a,l_b;x,t) $.
By solving Eq. (\ref{fourier}), we get the result
\begin{equation}
\widetilde{D} (0;0,0) = \frac{ \widetilde{D}^{(0)} (0,0) }
        {  1 - \lambda_2 (0) \widetilde{D}^{(0)} (0,0)  }
\end{equation}

The relevant dependence of $\widetilde{D} (0;0,0)$, as well
as of $\widetilde{D}^{(0)} (0,0)$, is on the temperature.
It becomes clear that, if the superconducting correlations
are such that $\widetilde{D}^{(0)} (0,0)$ has a divergence
at $T = 0$, then the effect of pair hopping gives rise to
the appearance of a pole at a finite value of $T$ in the
propagator of the Cooper pairs in the rope. According to
the conventional interpretation, 
this is the signal of the
condensation of Cooper pairs and the onset of the
superconducting transition in the system.

\begin{figure}[H]

\epsfig{file=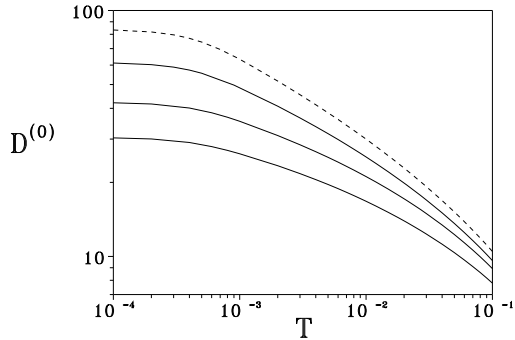, height=7cm, width=7cm,
         bbllx=0, bblly=0, bburx=600, bbury=600,
         angle=270}

\caption{Logarithmic plot of $\widetilde{D}^{(0)}$ at zero
frequency and momentum versus $T/E_c$. The solid lines are
the transposition of the solid curves in Fig. \ref{one} for
$n = 100$ and the respective values
$4g/(\pi v_F) = 0.75, 0.5, 0$ . The dashed line corresponds
to the case $n = 400$ and $4g/(\pi v_F) = 0.75$ .}
\label{three}
\end{figure}

In practice, when dealing with a rope of finite length $L$,
the divergence of $\widetilde{D}^{(0)} (0,0)$ is cut off at 
a temperature scale about one order of magnitude below 
$ v_F /L $. The curves shown in Figs. \ref{one} and 
\ref{three}, for instance, have been obtained for a 
system with $L = 1000/k_c $, which corresponds to $L \sim 1$ 
${\rm \mu m}$ with an appropriate choice of the length scale.

Taking our estimate of the pair-hopping amplitude
$\lambda_2 \sim 0.01$, we observe that the length $L =
1000/k_c $ may be in some samples at the limit below 
which a superconducting instability cannot arise in 
the system. This depends on the spatial distribution 
of metallic nanotubes, which should determine more precisely 
the effective value of $\lambda_2$ to be used. Quite 
remarkably, a superconducting transition has
been found in samples whose length is 1 ${\rm \mu m}$ or
greater, while a sample with low resistance and 
0.3 ${\rm \mu m}$ long has shown no 
transition at all\cite{sup}.

Relying also on our estimate for $\lambda_2$, we observe 
from Fig. \ref{three} that the temperature of the transition 
to the superconducting phase in a disordered rope with about
100 metallic nanotubes can be in the range between
$10^{-4} E_c$ and $10^{-3} E_c$. As the natural energy
scale in our model is $E_c \sim 0.1$ eV, this sets
the scale for typical transition temperatures in the range
between 0.1 K and 1 K . These values are consistent with 
the transition temperatures $T_c \approx $ 0.1 K and 
$T_c \approx $ 0.4 K measured experimentally in the two 
samples of Ref. \onlinecite{sup}.

The results of our analysis show that the superconductivity
in the ropes of nanotubes is close in nature to that of 
the alkali-doped fullerenes\cite{gun}. 
We have seen that the 
tunnelling of electrons between neighboring nanotubes in
a rope takes place in pairs,
which are formed within each nanotube due to the large
superconducting correlations which develop at low
temperatures.

The low values of $T_c$ compared to those of the alkali-doped 
fullerenes can be understood on phenomenological grounds 
by the fact that the electron-phonon coupling is smaller
in the nanotubes.
This is consistent with a higher 
estimate of $T_c$ obtained in Ref. \onlinecite{chi}
from measurements on small-diameter single-walled nanotubes,
which have a larger electron-phonon coupling than 
the nanotubes in the ropes. As seen in Fig. \ref{three},
another way to obtain a higher $T_c$ in the ropes
may be to increase the number of nanotubes of the 
samples. Finally, the feasibility of controlling the 
rate of tunnelling between the nanotubes should be 
studied since, as observed from Figs. \ref{one} and 
\ref{three}, a slight change in that parameter may
result in an increase of $T_c$ by more than one order
of magnitude.

Fruitful discussions with F. Guinea and A. Kasumov are 
gratefully acknowledged. This work has been partly 
supported by CICyT (Spain) and CAM (Madrid, Spain) 
through grants PB96/0875 and 07N/0045/98.


\begin{references}
\bibitem{rem}
For the sake of simplicity, we do not make explicit the
doubling of the linear branches at each Fermi point
characteristic of the carbon nanotubes.
\bibitem{rem2}
Although the points $l_a$ may not form in practice a regular
lattice, the assumption of a periodic arrangement in the 
transverse section of the rope
does not play an essential role in the
following considerations.
\end{references}
\end{document}